\documentclass{skaox2006}
\begin{document}
   \title{Cosmic Rays: Recent Progress and some Current Questions}

   \author{A. M. Hillas}

   \institute{School of Physics and Astronomy, University of Leeds,
             Leeds LS2 9JT, England}

   \abstract{A survey of progress in recent years suggests we are moving towards
a quantitative understanding of the whole cosmic ray spectrum, and that many
bumps due to different components can hide beneath a smooth total flux.
The knee is much better understood: the KASCADE observations indicate that the
spectrum does have a rather sharp rigidity cut-off, while theoretical developments
(strong magnetic field generation) indicate that supernova remnants (SNR)
of different types should indeed accelerate particles to practically this same
maximum rigidity.  X-ray and TeV observations of shell-type supernova remnants
produce evidence in favour of cosmic-ray origin in diffusive shock acceleration 
at the outer boundaries of SNR. There is some still disputed evidence that the
transition to extragalactic cosmic rays has already occurred just above
$10^{17}$ eV, in which case the shape of the whole spectrum may possibly be
well described by adding a single power-law source spectrum from many
extragalactic sources (that are capable of photodistintegrating all nuclei)
to the flux from SNRs.  At the very highest energy, the experiments using
fluorescence light to calibrate energy do not yet show any conflict with an expected
GZK ``termination''.  
(And, in ``version 2'',) Sources related to GRBs do not appear likely to play 
an important role.
}
   \maketitle
%

\section{Introductory overview\label{introsec}}

Because cosmic rays span such a huge range of energy, it is natural to 
start from a very deceptive broad view of the cosmic ray spectrum, such as that 
shown in figure 1, due to Gaisser (\cite{gaisserfig}), which shows the flux reaching the 
Earth, in the form of the energy carried by particles per unit interval of $ln(E)$, or 
$E^2 J(E)$, where $J(E)$ is the number of particles arriving per unit interval 
of time, area, solid angle and kinetic energy, E.  
\begin{figure}
\centering
\vspace{340pt}
\includegraphics{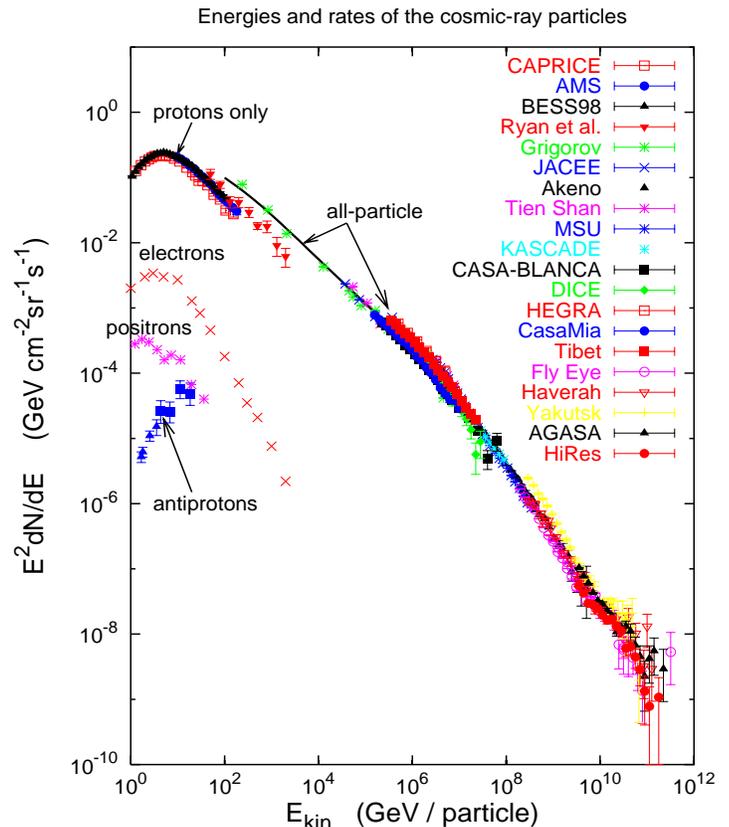}
\caption {Many measurements of the cosmic ray flux over a wide energy range,
assembled by Gaisser \label{gaisfig}}
\end{figure}
At the lowest energies, the fluxes of different nuclei can be measured, protons 
being the most numerous, and other common nuclei having practically the same shape 
of spectrum as a function of rigidity (momentum/charge $\propto$ energy/charge at these 
relativistic energies).  To identify the particles clearly, they have to be detected 
before they are broken up in the atmosphere, in detectors carried by balloons or 
satellites, and the flux is too low for this above about $10^5$ GeV ($10^{14}$ eV): 
beyond here the total flux of all particle types can be recorded by 
air shower experiments. 
The well-known power-law spectrum, $J(E) \propto E^{-2.7}$ holds to a good 
approximation before the ``knee'', the downward bend near $10^{15.5}$ eV, 
the fall-off below 
10 GeV being a very local effect within the solar system.  For 3 decades of energy 
above the knee the flux continues to fall somewhat more steeply, to the ``ankle'', 
where the rate of fall briefly becomes less steep again, until statistics
and possibly flux peter out near $10^{11}$ Gev ($10^{20}$ eV).  
At energies of several GeV there is good evidence from gamma rays produced in nuclear 
collisions (e.g. Hunter et al. \cite{huntergam}) that the cosmic rays originate 
in the Galaxy, 
and diffuse out; and the belief that the major source is acceleration at the outer 
shock boundaries of expanding supernova remnants (SNR) has strengthened recently 
in several ways, outlined below.

   It now seems likely that this bland shape masks a superposition of bumps and
variations which each tell their own story, though few of them can yet be disentangled
clearly, so this field of diagnosing the components is still very active.

     Recent experimental work at Karlsruhe (discussed in section \ref{galcompsec}) 
has made it seem very probable that the individual nuclear components each fall 
off rather steeply at a magnetic rigidity near $3\times 10^{15}$ V 
(i.e. at energies $3\times 10^{15}$ eV for protons, $6\times 10^{15}$ eV for 
helium nuclei, extending to $8\times 10^{16}$ eV for iron, the heaviest common nucleus).
Assuming this to be right, even though the point of turn-down for iron is just 
beyond the range of this experiment, the main Galactic component is made up of 
elemental components each extending to an energy near 
$Z\times 3\times 10^{15}$ eV (where $Z$e is nuclear charge), beyond where the 
fluxes turn down much more sharply than does the total flux that is plotted
in figure 1.  
Despite the separate bends, the total flux looks deceptively smooth after steepening 
a little at the knee, as shown in figure \ref{gaisfig}, 
at least as far as $10^{17}$ eV. 

      Beyond $10^{17}$ eV questions arise.  
The more extended gradual fall-off between the knee and the ankle has long 
been puzzling. 
Does some Galactic source (magnetars?) extend the spectrum of local particles 
well beyond $10^{17}$ eV (presumably highly-charged particles to allow them to 
be disoriented by Galactic magnetic fields)?  A widespread view had been that some
such additional component partly trapped within our galaxy eventually fell below 
the level of cosmic rays circulating throughout the universe, and originating in
rarer far more energetic sources.  The ``ankle'' might then mark the point where such
extragalactic cosmic rays became dominant, but this is not necessarily so, as will
appear from the discussion in section \ref{extragalsec}.

     One much older view of the knee-to-ankle region had been that the 
Galactic sources might accelerate particles to much higher energies than $10^{16}$ eV,
and the extended slightly steeper slope beyond the knee marked an increasingly 
rapid escape of particles from Galactic magnetic fields at higher energy, 
a view the present author mistakenly used when discussing anisotropies in a review 
22 years ago (\cite{araa}), when it appeared that there was an increasing anisotropy 
reflecting such a decreasing residence time.  
However, with much greater counting statistics, we now see no clear anisotropy 
apart from a small one in the region $10^{14}$ to  $10^{15}$ eV.  
Unless there is indeed a high-Z flux generated by Galactic magnetars, it now 
seems that the extragalactic component becomes very important at a much lower 
energy than previously thought.
If, then, there are no low-charge galactic particles above $10^{17}$ eV,
the failure to find convincing anisotropies would be explained. 
     These topics, and the particles of extreme energy, are discussed below.

     Cosmic-ray electrons will be mentioned only briefly.  
Figure \ref{gaisfig} shows that at a given energy they are much less numerous 
than protons --- 1--2\% around a GeV and even less at higher energies --- 
though their strong synchrotron radiation makes their presence in distant regions much 
easier to detect than that of protons and nuclei. 
The electrons may originate in SNR as we believe do the hadrons, or in plerions 
(e.g. the Crab Nebula), but if termination shocks of ultrarelativistic winds 
produce the acceleration in the latter, they probably accelerate an 
electron-positron medium, and the low relative abundance of cosmic-ray positrons 
indicates that plerions do not form a major source.

\section{The upper end of the main Galactic component of cosmic rays 
  \label{galcompsec}}

     Figure 2 shows an enlarged view of the spectrum from 
$10^{11}$ to $10^{19}$ eV, by plotting $E^{2.75} J(E)$.
\begin{figure*}
\centering
\vspace{300pt}
\includegraphics{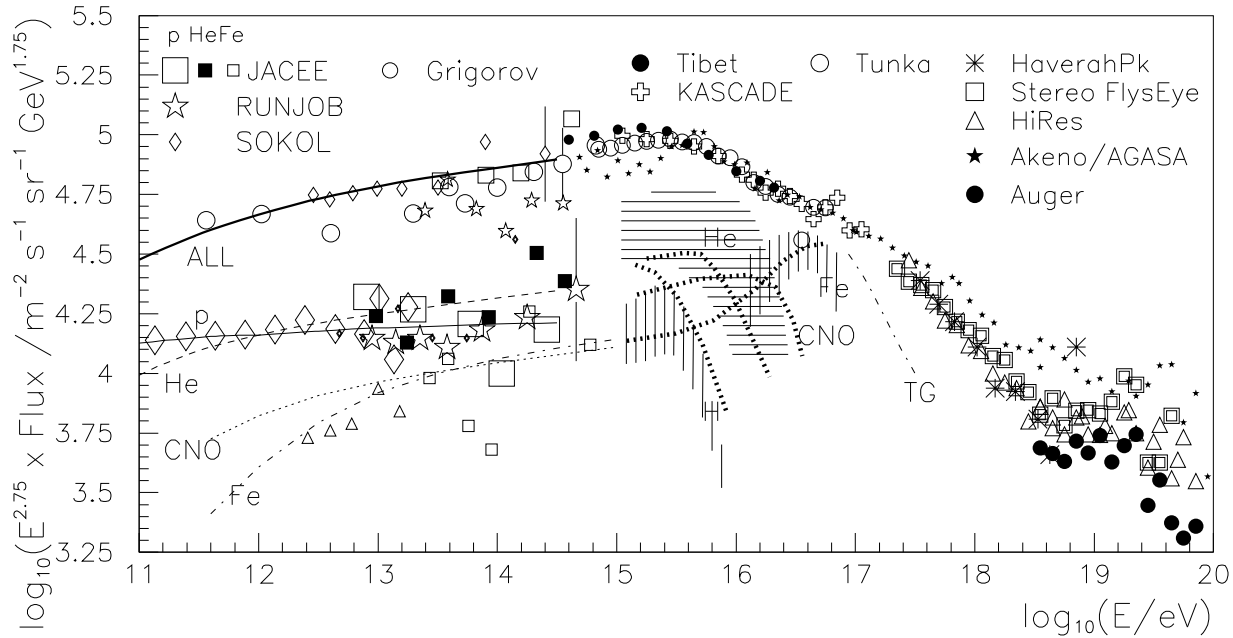}
\caption {Showing the well-defined shape of cosmic ray energy spectrum above 
$10^{15}$ eV derived from air shower experiments using several different 
approaches to energy measurement, forming a continuous extension of the 
spectrum obtained from (mainly) balloon-borne experiments at $10^{11} -- 10^{15}$ eV.  
At the latter energies, spectra of some individual nuclear groups are shown 
by lines (He may be a little too high), with a few data points for 
p, He (filled points) and Fe.  
Above $10^{15}$ eV, the small circles (p,He), stars (CNO) and triangles (Fe) 
show the provisional decomposition of the flux into 4 nuclear 
groups by KASCADE \cite{haungs}.
   \label{spec1}
 }
\end{figure*}
Above $5\times 10^{14}$ eV, the plotted data points come from mature air shower 
experiments selected to represent several different techniques used for obtaining 
the shower energy --- density of particles at the ground, 
number of particles near shower maximum, Cherenkov or fluorescent light measuring 
the energy deposition in the atmosphere, or very detailed measurements of many 
particle components --- and all using very similar quark-gluon-string (QGS) hadronic 
interaction models to make detailed predictions of the observed shower parameters.  
(References to the experiments in Hillas \cite{hillasdsar}, \cite{hillascris})   
Up to $10^{18}$ eV, such selected experiments are seen to be in very good agreement, 
though there are two high-exposure experiments at the highest energies which revealed 
discrepancies still to be understood, and which the Auger project was designed to clarify: 
the Akeno-AGASA spectrum seems to drift gradually above the others, and the Yakutsk 
spectrum seems to assign systematically higher energies.  
The region above $3\times 10^{18}$ eV will be discussed in section \ref{extragalsec}.
At energies below the knee, some measurements of proton, helium, CNO and iron nuclei 
made above the atmosphere are shown, though most helium points 
are omitted for clarity because of wide scatter.

The point of this diagram is to show that air shower measurements fit well on to
the direct measurements made above the atmosphere, so we can try to understand
the whole spectrum in detail.  In particular, the slight steepening of the spectrum
from knee to ankle was not understood.  One anticipated that accelerators would be
able to accelerate each type of ion to the same magnetic rigidity, indeed with
very similar rigidity spectra, except insofar as different accelerators (SNR)
perhaps had expanded into gas of different composition.

      In the air shower domain, where direct identification of the primary particles is 
not possible, two methods of making rough estimates of the primary particle's mass have 
been attempted for a long time: 
(a) estimation of $x_{max}$, the depth ($x$ in $\mathrm{g\ cm^{-2}}$) in the 
atmosphere at which the shower reaches its maximum number of particles; or 
(b) measurement of the proportion of muons in the shower when it reaches the 
observing level (``$N_\mu /N_e$'').  
Both depend on the fact that with a primary particle of energy $E$, containing 
$A$ nucleons, each nucleon to a first approximation makes its own shower of energy 
$E/A$ (coaxial with the others); and that the number of shower multiplication 
steps taken (and the mass of air traversed) for the particles to fall to the 
critical energy, at which multiplication ceases, increases logarithmically with 
($E/A$), and hence falls with $A$, for showers of a given $E$.  
The backbone of the shower is a hadron cascade, in each step of which 1/3 of the 
pions decay instantly to drain off energy into the electromagnetic shower, 
so the more cascade steps that are required before the charged pion energies 
get down to tens of GeV, at which decay to muons can occur, (i.e. the higher is 
$E/A$), the less is the fraction of energy left for muon production.
  
Both approaches have provided unclear indications that the mean mass $A$ starts to 
rise beyond the knee, but the observable effects are not large, a situation which is not
surprising if the mean mass changes gradually, as one element after another becomes
less prominent.
In the present author's opinion, a notable advance has been made in recent years, after
several decades of inconclusive observations in this difficult area.
If one wants to do better in elucidating the composition, 
and separate the showers of different primary mass, 
$A$, the $N_\mu /N_e$ approach would give better resolution (separation of 
$N_\mu /N_e$ curves for different mass groups, relative to shower-to-shower 
fluctuation), than the $x_{max}$ --- but only if a very extensive area of muon 
detectors is available to make statistically excellent $N_\mu$ measurements in
individual showers.  
In the KASCADE array at Karlsruhe, a very densely packed array of electron and 
muon detectors was set up for this purpose, and figure \ref{spec1} 
shows how their attempted analysis of the cosmic rays into mass groups lines up 
with the direct measurements at lower energies.  
Despite the results looking somewhat confusing, they are the best we have, 
and deserve some scrutiny.

     The dotted lines starting at $10^{15}$ eV indicate the 
original preliminary unfolding of the spectrum into components H, He, CNO and Fe 
(Haungs \cite{haungs}), 
and are shown only to demonstrate the nature of the pattern, with individual knee 
energies consistent with a $Z$ proportionality --- a magnetic effect which could be
related to the dimensions of the accelerator.  
The KASCADE workers would not wish any emphasis to be placed on this earlier version, 
but it gives an orientation to the reader not very familiar with this field, 
as their later analysis is harder to assimilate.  
In their more detailed attempt to unfold the spectrum into even more components 
(Antoni et al. \cite{kascadefin})
the results became somewhat unstable in the centre of the mass range, since the 
$N_\mu /N_e$ distributions for different mass groups overlap greatly, and where 
changing from a QGSJET hadron interaction model to a SIBYLL model changed the 
predicted $N_\mu /N_e$ a little.  
The upper and lower edges of the hatched regions in figure \ref{spec1} correspond 
to the flux deduced to correspond to H (or He or Si-Fe) when using one or other of 
these models.  In the particular case of the SIBYLL curves, the ``effective maximum 
energy'' for particular nuclei may be taken, say, as the energy at which the flux 
$E^{2.75}J$ has fallen by half a decade from its peak value, near its knee, and is 
at about 6, 11, 34 and $>100 \times 10^{15}$ eV for H, He, CNO, and Fe --- 
quite consistent with proportionality to Z (=1, 2, 7, 26).  
For clarity, the very uncertain CNO fluxes are not shown: 
at $10^{16}$ eV, the slight model differences cause SIBYLL to 
attribute the majority of the flux to CNO, whilst QGSJET labels it as He.  
So one cannot yet distinguish all these nuclear groups in the middle of the range; 
but the interesting features are that 
(a) the individual elemental fluxes appear probably to fall sharply above a certain 
magnetic rigidity --- E near $Z\times 3\times 10^{15}$ eV) --- though only separated 
clearly for the case of hydrogen; 
(b) even before $10^{15}$ eV, the proton flux may 
indeed have been falling a little more steeply than the others, so that protons do 
not dominate at the knee; and 
(c) the flux of the heaviest nuclei --- Si-Fe, 
and possibly CNO --- may rise distinctly immediately before the knee.  
This rise may largely be an artefact of the analysis (and difficulty in including
Ne-Mg components), though it 
might be a first indication of the additional 
acceleration and release of newly-synthesised elements from the exploding star
itself, in the very early life of 
the SNR --- perhaps injected at the inner shock.

     These elemental components add up to an all-particle flux matching the data 
points as far as $10^{17}$ eV, but then falling steeply, somewhat as shown by the 
line TG (total Galactic) if the iron does indeed have a steep rigidity turn-down 
like the lighter elements.  

In summary, there is reasonably good initial evidence that the spectra of 
individual nuclear components do turn down much more sharply than would be guessed
from the slightly-bent knee of figure \ref{gaisfig}.
Why does the superimposed output from many SNR of different mass and in different 
environments yield a distinct ``maximum'' energy?
And why does the flux not drop more steeply above $10^{17}$ eV, like the line TG?
Does the overall Galactic component indeed turn down rather sharply here?
Evidence for a change in the nature of the particles above $10^{17}$ eV, through
further well-calibrated measurements of shower properties here will be welcome. 

Before describing work on the high-energy extension of the spectrum, developments 
in theoretical understanding of the source of Galactic cosmic rays and the
distinct maximum near $3\times 10^{15}$ eV, and relevant 
new observations of SNR will be discussed.

\section{Further support, and questions, for the model of origin of Galactic 
cosmic rays at the boundaries of expanding SNR \label{shockacsec}}


\begin{figure}
\centering
\vspace{200pt}
\includegraphics{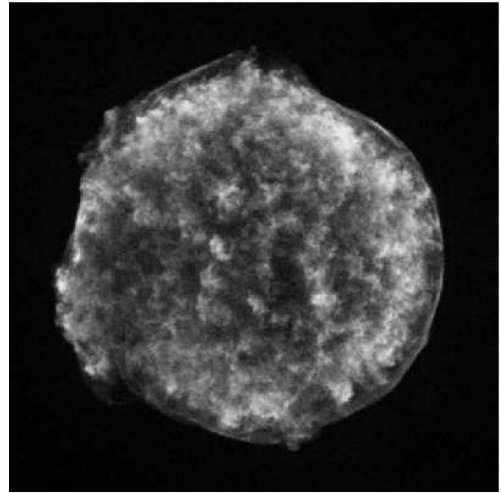}
\caption {2005 CHANDRA image of Tycho's SNR, from Chandra.nasa.gov.
The outer thin surface is synchrotron radiation from highly relativistic electrons
accelerated at the outer shock.  Behind this is a highly turbulent region, 
presumably formed by Rayleigh-Taylor instability at the contact discontinuity.
Acknowledgement to NASA/CXC/SAO.
 \label{snchan1}
}
\end{figure}
\begin{figure}
\centering
\vspace{200pt}
\includegraphics{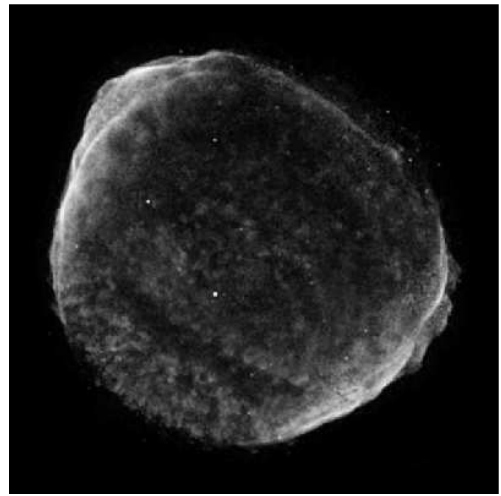}
\caption {CHANDRA image of SNR1006, from Chandra.nasa.gov.
Synchrotron radiation from ultrarelativistic electrons occurs in two
'polar cap' regions, probably where the external magnetic field is nearly
perpendicular to the outer surface.
Acknowledgement to NASA/CXC/SAO.
 \label{snchan2}
}
\end{figure}

Radio telescopes have for long drawn attention to supernova remnants (SNR) as 
principal sources of multi-GeV electrons, through their synchrotron radiation, 
but high-energy X-ray images of supernova remnants, especially by 
CHANDRA \cite{chandraweb}, have revealed fresh detail.  
Figures \ref{snchan1} and \ref{snchan2} show an extremely narrow smooth shell of 
synchrotron radiation at the outer edge of SN1006 and Tycho's SN, outside the thermal 
heavy-element gas.  
The extreme thinness is consistent with the very short radiative cooling lifetime 
of extremely relativistic electrons ($\sim 10^{13}$ eV) in magnetic fields of 
several hundred microgauss (V\"olk, Berezhko \& Ksenofontov \cite{volkhimag},
 Ballet \cite{balletxsyn}) --- very persuasive evidence that this 
SNR outer boundary is indeed where the relativistic electrons do gain their energy, 
rather than, for instance, in the highly turbulent region generated from 
the contact discontinuity, clearly seen in the Tycho image, which was once 
considered as a possible site of acceleration.  
In Cas A a very thin outer emitting layer is also seen, but is much less smooth, 
evidently because of the disturbance of the expansion while passing through a 
shell of wind-compressed gas that was shattered en route 
(leaving behind the famous clumpy knots).

    The model of diffusive shock acceleration at the boundaries of SNR provides 
much the most persuasive theoretical basis for interpreting Galactic cosmic rays, 
but as the complex details of particle-plasma interactions are far from fully 
understood yet, an optimistic simple-minded approach will be taken, to assess progress.
The main features of this model are that, to a first approximation at least, 
it naturally 
(a) produces an extended power-law energy spectrum of about the right slope, 
(b) is such an efficient energy converter that it is hard to find a convincing
rival model, 
(c) it explains several features of the elemental composition of cosmic rays, and 
(d) now seems likely to explain a distinct spectral knee (for SNR sources) 
at the observed position, which was not the case a few years ago.  
Consideration of the relationship between these basic aspects will reveal some of 
the remaining questions.  The features (a) to (d) will be discussed in turn.

 (a) At an over-simple sharp shock, where no pressure can be transmitted ahead of the 
shock, the density jumps by a factor 4, there is correspondingly a sudden 
change in bulk gas velocity across the shock, and some individual particles 
diffusing in the gas can cross back-and-forth, gaining energy by the 
well-known Fermi process of bouncing between two relatively convergent gas masses.  
The resulting momentum spectrum behind the shock is $dn/dp \propto p^{-\gamma}$, 
where $\gamma$ is between 2, for highly supersonic flow, and about 3 when the 
speed has slowed down to a mildly supersonic level.  
In a SNR, most of these particles remain trapped (new ones being constantly 
accelerated while those previously accelerated are gradually adiabatically 
decelerated by 
the expansion), until some stage when the old remnant condenses, breaks up, and 
releases them.  
It is unclear just what is the spectral exponent at the break-up stage.  
To produce Galactic cosmic rays with a spectrum $E^{-2.7}$, as seen over 
most of their energy range, one would need $\gamma \approx 2.37$ if their 
residence time in the Galaxy $\propto E^{-1/3}$, as for Kolmogorov scattering, 
which most probably governs their propagation.  
(Nuclear spallation studies at low energies gave an apparent $E^{-0.6}$ 
dependence of residence time within the Galactic gas, but this was distorted 
by effects of moderate energy changes during scattering, and would in any case 
have led to abnormally large anisotropy near $10^{16}$ eV.)  
The variation of the spectrum within a SNR up to the time of release still 
needs a careful study to check that this is reasonable.  
Inside a young SNR, $\gamma$ would be close to 2.  
However, the cosmic rays themselves must broaden the shock, as shown below, 
so that numerical studies of particle acceleration in SNR have predicted a departure
from a simple power law, with a steeper slope at very low energies ($\sim$GeV), 
and a flatter slope ($\gamma \sim 1.8$) within 1 or 2 decades of the maximum energy. 
There is as yet no sign of such a general upturn in the cosmic-ray spectrum, so
we must look to TeV telescope arrays such as HESS, VERITAS and MAGIC to seek
this upturn in young SNRs such as Tycho.

\begin{figure}
\centering
\vspace{120pt}
\includegraphics{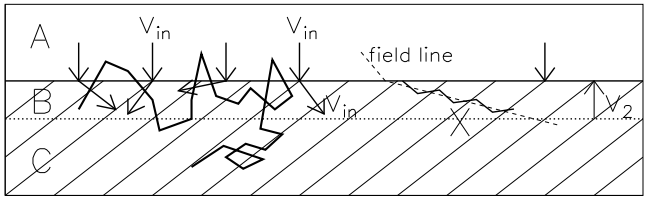}
\caption {Schematic diagram of gas crossing a collisionless shock, seen from the
frame in which the compressed gas C is at rest.  
External gas A enters with speed $v_{in}$: as each new layer B is added, the position
of the shock moves up with speed $v_2<v_{in}$.
\label{vinshock}
 }
\end{figure}

 (b) One needs a very efficient accelerator for cosmic rays, for the rate of 
supply of cosmic ray energy in the Galaxy needed to compensate for leakage out is 
$\approx 1.5\times 10^{34}$ W, or about 1/6 of the $1.0 \times 10^{51}$ erg of 
kinetic energy (excluding neutrinos) supplied 3 times per century by exploding 
supernovae (see e.g. Hillas \cite{hillasdsar}).  
(The assumption of 3 SN per century is perhaps the most uncertain number 
involved here.)  
The underlying reason why this process is so efficient is illustrated in 
figure \ref{vinshock}, a schematic section through a small part of the shock, 
seen in the rest frame of the hot compressed gas C behind the shock.  
Unshocked gas enters at speed $v_{in}$ from the external medium A, and is stopped 
(in this frame) by the pressure of the hot gas behind the shock.  
But, in this frame, this pressure does no work, so the bulk motion stops, but the 
incoming particles retain their kinetic energy and their speed $v_{in}$, 
though their directions are randomised in a very narrow transition layer.  
If the layer B of compressed gas has been added in the last second, the shock 
has receded by this distance: it has a recession velocity $v_2$ upwards.  
However, the particles are moving around with speed $v_{in} = (\sigma - 1) v_2$, 
where $\sigma$ is the compression ratio across the shock: 
the particle speed is thus considerably greater than the shock recession speed $v_2$, 
and a large proportion of the particles can readily scatter (diffuse) back 
to re-cross the shock (thick random-walk line in the diagram) and 
start on the repeated ``bouncing'' process of Fermi acceleration.  
Potentially, most particles can get injected into the 
acceleration process, and steal all the available energy.  
Presumably the thin transition layer will have to adjust until there is no 
catastrophe.  
A considerable reduction of the gas temperature will be required, and as a crude 
representation of this one may expect that about half the internal energy of the 
shock-heated gas is taken by the ``cosmic rays''.  
(It seems that this should rightly be referred to as ``the other Fermi gas'', 
having a modification to the classical thermal momemtum distribution at high 
momenta quite different from the truncation seen in the well-known 
quantum ``Fermi gas'', 
but of a form also deriving from Fermi.)
Even if half the energy is in the Fermi tail, this may represent only 
$10^{-3}$ to $10^{-4}$ of the particles.   
Then, as a result of the accompanying adiabatic expansion of the SNR gas, 
one would expect the SNR's energy to be divided about equally between bulk kinetic
energy of outflow, ordinary thermal energy and relativistic ``cosmic-ray energy'', 
for a long time during the evolution.  
More detailed simulations by Berezhko, Ksenofontov and collaborators
(e.g. Berezhko et al. \cite{bksims}) are roughly in line with this.
\begin{figure}
\centering
\vspace{140pt}
\includegraphics{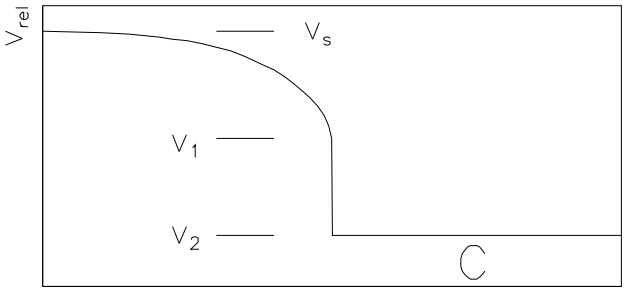}
\caption {Speed relative to shock as gas approaches and crosses the shock.
Gas flows from left: C is the compressed hot gas behind the shock; $v_s$ is the
speed of the shock relative to the undisturbed external gas.
In this frame, the shock is stationary, the pattern is quasi-steady, and the product 
$density \times v$ is constant.
This shock precursor is formed as ``bouncing'' (diffusing) relativistic
particles push against and take energy from the incoming gas.
Very energetic diffusing particles with high Larmor radius wander
further ahead of the shock, and recoil from higher-speed apptroaching gas, 
receiving a larger fractional energy gain per crossing than do most of the particles.
\label{precurspic}
 }
\end{figure}

    However, there is a question of suppression of injection. Where magnetic field 
lines (inside the SNR) cross the shock front at a large angle to the shock normal, 
ions guided by the field will be constrained to move towards the front at a rate 
less than $v_{in}$ (example X in figure \ref{vinshock}), and the re-crossing process 
could never start.  
V\"olk \cite{volkpocap} uses an estimate that ~20\% of the front would be an active accelerator 
for this reason.  
The ASCA and CHANDRA images of SN1006 (figure \ref{snchan2}) may indeed exemplify this.  
However, Tycho and other SNR show activity over a large proportion of the front 
(perhaps where the field has become more turbulent).  
If, then, one crudely takes the average proportion of the SNR surface over which 
injection is not magnetically suppressed to be about 50\%, one is left with 
1/6 of the SNR energy being present in cosmic rays, which is in good agreement 
with what is required to account for Galactic cosmic rays.

    But the cosmic rays around the shock (figure \ref{vinshock}) have gained 
their energy by bouncing from the incoming gas (zone A).  
They have taken their energy from this inflow, hence slowing it down just 
before it reaches the shock.  
Figure \ref{precurspic} shows schematically the gas velocity plotted against 
distance from the shock --- this time, velocity relative to the shock.  
As a consequence of this ``precursor'' structure, the most energetic cosmic rays, 
which diffuse further around the shock during their ``bouncing'', experience a 
greater velocity jump, and gain more energy at each excursion; so the spectral 
exponent $\gamma$ should be less than 2 (say, 1.8) for the upper 
2 energy decades or so (see e.g. V\"olk et al. \cite{tychospec}).  
This seems an inevitable consequence of efficient acceleration, and it now seems 
surprising that the observed cosmic ray spectrum is as ``straight'' as is 
generally observed! One consequence of cosmic rays stealing a large fraction of the 
internal energy of the hot gas is that the usual hydrodynamic models of SNR 
will be inaccurate, especially regarding the gas temperature
(Ellison et al. \cite{ellisonhydro}).  
It has been noted that the compressed shell shown in the CHANDRA image of 
Tycho is thinner than expected from a hydrodynamic approach omitting cosmic 
rays (e.g. Warren et al. \cite{warrentycho}).

(c) The elemental composition of the accelerated particles should be that of the 
gas swept up by the outer shock.  In figure \ref{vinshock} these ions all enter 
at the same speed, $v_{in}$, and the Fermi acceleration builds up a 
power-law momentum spectrum starting from this initial velocity: 
the result is that the different types of accelerated particle should have the 
same proportions as in the gas if measured at the same velocity or Lorentz 
factor --- i.e. at the same $E/A$, as observed --- a trade mark of the 
shock injection.  
Observed preferences for certain elements have been modelled as effects of 
the modified shock structure (figure \ref{precurspic}) which gives some 
increased energy gain to particles of higher mass-to-charge ratio as they 
penetrate a little further ahead of the shock while bouncing, especially for
those ions which on entry are 
initially bound inside charged interstellar grains 
(Meyer et al. \cite{meyerdust} and Ellison et al. \cite{ellisoncomp}). 
(This model re-interprets abundances supposedly related to an atom's first 
ionization potential as instead reflecting its condensability into grains.)  
It has become standard doctrine that atoms from the external medium, 
and not the supernova ejecta, are thus accelerated, and the explanation of 
the elemental composition is one of the main strengths of the shock acceleration model.  
However, during the early phase of the SNR expansion, the ejected SN material 
sweeps into the compressed shell through its inner shock, and there seems to be
no reason why its ions should not be accelerated there.  
Normally, this body of accelerated supernova gas would then suffer great 
adiabatic cooling during the huge expansion of the SNR before release, and so 
would make a very small contribution to cosmic rays, but there may perhaps be an 
exception at the knee, mentioned in paragraph (e), below.

 (d)  It was long believed that diffusive shock acceleration would yield a maximum energy
 near $Z \times 10^{13}$ eV (Lagage and Cesarsky \cite{lagces}) --- or a factor 10 higher
 if one overruled Lagage and Cesarsky's objections to Bohm scattering in the internal
 magnetic fields.  
Lucek and Bell (\cite{lucbell}), however, have reported numerical 
simulations of interactions between ions streaming in a magnetic field 
(as happens where cosmic rays 
diffuse just ahead of a shock) and the plasma carrying the magnetic field, showing that 
Alfv\'en-like magnetic disturbances grew very rapidly to magnetic amplitudes much greater 
than the initial magnetic field, and Bell and Lucek (\cite{belluc}) went on to consider 
the likely magnitude of highly contorted magnetic field that would be generated just 
ahead of the SNR shock by the particles being accelerated.  
Their result can be expressed in the following formula (S.I.units) for the r.m.s. 
magnetic field strength $b$ arising only from Fourier components within one 
e-fold range around the wavelength resonating with particles of any particular energy
\begin{equation} b = V_{shock}\sqrt{\mu_0 \rho_{gas}} (\eta/20),
\end{equation}
 where $\eta=P_{cr}/(0.1 \rho_{gas} V_{shock}^2)$.
Here $V_{shock}$ is the velocity of the shock advancing into a medium of density 
$\rho_{gas}$, and $P_{cr}$ is the pressure of accelerated particles in one e-fold range.  
The quantity $\eta \sim 1$ in an efficient shock. (Although this magnetic energy is
proportional to the ram pressure, it is less than equipartition strength.) 

This $V_{shock} \sqrt{\rho_{gas}}$ dependence will change the systematics of the maximum 
energy of the spectrum of accelerated particles in SNR.  
Firstly, $E_{max}$ is attained 
very early, well before the sweep-up time ``$T_0$'', and the most energetic particles 
then very soon escape, as the field weakens.  
Secondly, the $\rho_{gas}^{1/2}$ dependence of $B$ introduces a similar factor
into the rate of gain of energy, which almost cancels a dependence 
$\rho_{gas}^{-1/3}$, coming from the time available for energy gain, 
found in previous models for $E_{max}$ where $B$ was constant;
and Hillas (\cite{hillasdsar}) found with a ``toy model'' for spectrum generation, 
that several different kinds of SNR placed in different external environments 
generated spectra terminating in sharp downward bends all very close to 
$2 \times 10^{15}$ V rigidity.  
Other theoretical studies have examined the effect of various other 
wave-field-damping processes and
nonlinear wave interactions, to expand the treatment of Bell and Lucek.
Ptuskin and Zirakashvili (\cite{ptuszir}) quote model energy spectra with a 
knee near the same place, and remark on the very steep fall above the knee. 
Marcowith et al. (\cite{marcow}) 
expect to get a maximum energy near the observed knee, and also expect to get
a cosmic-ray spectral index $\gamma$ steepened to $\approx 2.3$ (for explosions into the
warm phase of the ISM) because of energy losses by the bouncing particles when forming
the strong magnetohydrodynamic turbulence.  Bell did not expect this effect to be
significant, though.  
Clearly there is much to do before we understand sufficiently fully these strong ion-plasma
interactions, but if the prescription of Bell and Lucek is about right, one may well be 
able to account for a distinct knee in the superimposed spectra of many different 
SNR --- virtually at the observed position.   
There is some observational evidence that the magnetic field in SNR has 
reached levels of several hundred microgauss, some of it from the rapidity of 
the cooling of relativistic electrons behind the front 
(V\"olk et al. \cite{volkhimag}, Ballet \cite{balletxsyn}, Vink).

 (e) If it is true that the most energetic
 particles are generated early, especially when a fast shock ploughs through a 
dense fossil stellar wind, and these particles are soon released as the magnetic field 
weakens, it seems possible that heavy elements from the supernova ejecta may be 
accelerated to knee-energies at the inner shock, and allowed out very early,
before they are adiabatically cooled.  
This may perhaps make some contribution to the Si-Fe component in the 
KASCADE elemental unfolding shown in figure \ref{spec1}.

The beautiful CHANDRA pictures show only radiation by relativistic electrons, 
and are blind to protons.  
We now look to TeV gamma rays to show us the hadrons in SNR., when doubts about
the TeV contribution from inverse-compton scattering of electrons are clarified.
Up to now, excellent images from HESS reveal striking shells in cases of 
complex dense environment that provide a massive target for the cosmic rays to 
interact with, but are hard to analyse quantitatively 
(especially RX J1713.7--3946, Aharonian et al.\cite{aharrjx}).
However, these TeV gamma rays are most probably due to hadrons rather than electrons 
(\cite{aharrjx2}), and if $B$ in SNR is indeed larger than previously thought,
there must be fewer electrons present to make the observed synchrotron radiation,
so the problem of electron background in TeV SNR images may be receding. 
The instrumental sensitivity is still not adequate to study SN1006, 
a very clean system but in a region of very low gas density.  
Before long, Tycho's and Kepler's SNRs and Cas A should be measurable.

This has been an optimistic assessment of the SNR scheme for cosmic ray 
generation, but observational challenges should be taken seriously.  
Plaga (\cite{plaga}), for instance, has emphasised two problem areas.  
Firstly, the absolute TeV luminosity of certain SNRs or SNR features is well 
below what has been reasonably expected if they are efficient hadron 
accelerators; and secondly, the Galactic distribution of cosmic-ray protons 
has not appeared compatible with a much more centrally concentrated supernova 
distribution.  As regards the first point, we should soon have more sensitive 
measurements to examine.  
As regards the second, I believe that the deduction of cosmic-ray radial 
distribution beyond $\sim 10$ kpc from the Galactic Centre from gamma ray maps 
is, so far, inherently wildly inaccurate, as one can see by comparing the
very different distributions obtained by different methods (e.g. Strong and
Mattox, \cite{smatx}, Bloemen, \cite{bloe}, Hunter et al., \cite{huntergam}).

\section{Extragalactic cosmic rays \label{extragalsec}}

     Above $10^{19}$ eV the magnetic field in the vicinity of the Galaxy would not 
trap very effectively even the very heaviest nuclei; but there is still little 
clear evidence of anisotropy, so the flux here is presumed to originate much 
further away, most probably in systems dominated by jets from active galactic nuclei
in order to provide the necessary exceptional energy source.  
In this case, the flux is expected to drop sharply above $5\times 10^{19}$ eV 
(the GZK effect) due to energy losses accompanying pion production reactions 
between nucleons and cosmic microwave background photons (CMBR), which would 
greatly attenuate the flux originating much beyond the local Virgo supercluster 
(depending on the exact energy).  If the sources were stronger in the past, 
behaving somewhat like the rate of massive star formation (as for gamma ray 
bursts), there would also be a (less severe) flux drop near $5\times 10^{17}$ eV, 
due to pair-production by CMBR on protons (or nuclei at a higher energy) as 
distant regions of the universe would dominate at energies below the
threshold for energy losses --- in effect an earlier part of the GZK fall.  

At the highest energies, beyond the ankle, the cosmic ray fluxes reported by 
different experiments did not agree so well.  Here, widely spaced detector arrays 
sampling the outer fringes of showers had to be used until the technique of 
measuring the fluorescent light emitted by very distant showers gradually became
well developed.  A few very extended arrays of shower detectors had failed to find 
evidence of the expected fall near $5\times 10^{19}$ eV (see the AGASA points in 
Figure \ref{spec1}), stirring up hypotheses concerning decay of long-lived 
super-massive particles supposedly clustering near the Galaxy, 
and generating secondaries of $\sim 10^{20}$ eV, 
or failure of normal relativistic energy relationships in collisions at extreme 
Lorentz factors.  
The Auger project was established to clarify the spectrum at these extreme 
energies --- using both particle detectors on the ground and recording fluorescent 
light in the atmosphere to check the shower energies --- and to record 
enough cosmic rays above $8\times 10^{19}$ eV (where energy loss rates should
make the distances to detectable sources small) to see which exceptional
objects they were coming from.

\begin{figure}
\centering
\vspace{210pt}
\includegraphics{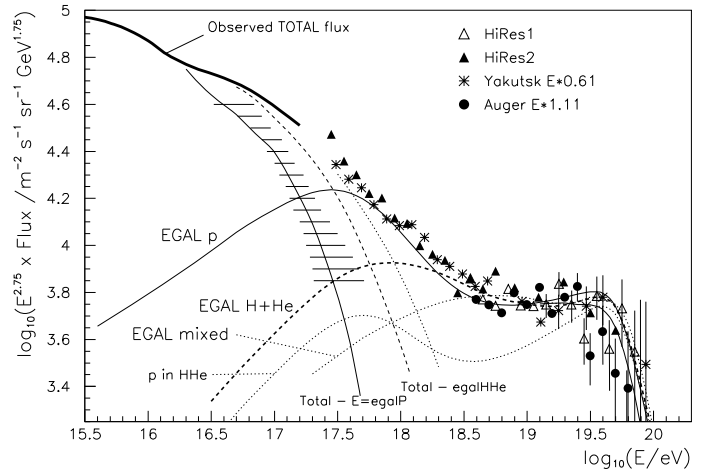}
\caption {The upper end of the cosmic ray spectrum. Data from HiRes (as
in Figure 2), from preliminary Auger exposure (energy scaled by factor
1.11 to match HiRes), and Yakutsk (energy scaled by 0.61).
Thick line shows mean of data points below $10^{17}$ eV from fig. 2, and
hatched area shows probable fall-off of Galactic flux if the rigidity spectrum
of Si-Fe component falls off as steeply as other components appear to do
in the KASCADE spectra.
The thin line `EGAL p' shows the expected flux of cosmic rays from
universal sources accelerating only protons, with spectral exponent 2.4,
and evolving in time like star formation rate, subject to energy losses
en route (CMBR and estimated starlight-infrared interactions).
The lower branch above $10^{19}$ eV shows the effect of terminating the
spectrum at source near $5\times 10^{20}$ eV, rather than at $10^{22}$ eV.
`EGAL H+He' likewise, but assumes the sources accelerate a primordial mix 
of H and He, with spectral exponent 2.2.
`EGAL mixed' is from Allard et al., taking a normal composition, exponent 2.3
and unvarying source strength.
 `Total-egal' curves show the flux required from other sources (presumably 
Galactic) to make up the observed total.
The data suggest that the flux may be falling even before the
expected GZK drop (seen in the curves near $5\times 10^{19}$ eV), presumably
due to accelerators' maximum energy.
\label{augere}}
\end{figure}

    The spectrum reported by the partly-built Auger array (Sommers \cite{augerspec}) 
is only preliminary, and results from a short initial exposure.  It agrees quite 
well in shape with the last of the previous fluorescence detectors, 
HiRes (Thomson \cite{hiresspec}),
and with the shape of the spectrum from the very early large array at Yakutsk
(Egorova et al. \cite{yakspec}).  These three spectra are shown in 
figure \ref{augere}, with the energy scales adjusted to bring the fluxes 
($dN/dlnE$) into alignment below $10^{19}$ eV, so that they fit onto the several 
concordant experiments shown in figure \ref{spec1}.  
To achieve this, energies were multiplied by the factors 1.11 and 0.64 for 
Auger and Yakutsk respectively.

     These measurements appear to be compatible with the expected GZK fall 
near $5\times 10^{19}$ eV, although this should become much better defined 
after another 2 years of exposure.  Their shapes are also suggestive of the effect 
of the earlier CMBR-pair-production-induced fall near $5\times 10^{17}$ eV.

Supposing that diffusive shock acceleration again converts a huge available gas 
kinetic energy into an extended power-law spectrum of particles made up of 
the locally available ions, one can calculate the characteristic large 
changes imposed on the spectrum by interactions mainly with the CMBR 
while travelling to the Earth, and compare with the observed cosmic ray spectrum.
There are several free parameters available to fit the data on spectral shape 
---  spectral energy exponent, and maximum energy, $E_{max}$, at the source, 
source composition, and variation of source power with 
cosmic time, even without supposing the sources are unevenly distributed in 
space (which is significant only above $\sim 4\times 10^{19}$ eV
or below $\sim 10^{16}$ eV) --- and 
of course there could be different types of source.  
In view of this, the consequences
of just the simplest natural assumptions will be illustrated, without
optimising the fits through wide-ranging variations.

     If we adopt the most plausible assumption that these very powerful 
high-energy extragalactic sources depend on shock acceleration powered by AGN jets ,
we still do not know how far from the AGN the shocks form, 
and what is the source of the ions swept into the shock.  
If the matter has passed close to the AGN after having gained an initial 
energy boost, it might be pure hydrogen as a result of photodisintegration 
of other nuclei at source.  
If acceleration occurred near the termination of 
the jet, at the typical large radio lobes, the injected matter might be 
H and He, similar to primordial composition.  
Or its composition might be not unlike galactic matter.  
Starting with the same rigidity spectrum for each type of ion, the spectrum 
at Earth would be changed by photodisintegration and energy losses en route, 
due to the CMBR and light from stars, mostly infrared.    
Illustrated in figure \ref{augere} are the simplest fits based on different 
types of source composition just mentioned.  

   (a) `EGAL p' is a pure proton source,  
$J_{source} = k.E^{-2.4} exp(-E/E_{max})$, and a source strength evolving 
like the star formation rate (Porciani and Madau SFR2 \cite{porcimad}).  
The higher curve (above $3\times 10^{19}$ eV) has 
$E_{max}$ set at $10^{22}$ eV, the lower at $5\times 10^{20}$ eV: 
it makes little difference below $10^{20}$ eV.  
(Note that below $10^{17}$ eV the flux gives the appearance of rising with 
increasing energy  merely because it has been multiplied by $E^{2.75}$ 
for plotting.)  
This differs a little from the form advocated by Berezinsky, Gazizov and 
Grigorieva (\cite{berez1}), essentially in assuming 
the time-evolving source power.  
They assumed a constant source (which is certainly a possibility for such jets), 
which necessitated a more steeply falling spectrum  ($\sim E^{-2.6}$ to $E^{-2.7}$). 
But with stronger sources in the past, there are more long trajectories to suffer 
energy loss when pair-production sets in above $5\times 10^{17}$ eV: 
without this evolution in production rate, a more steeply-falling source 
spectrum is needed to give the same slope seen in the data in this region.  
(With their steeper spectrum, Berezinsky et al. introduce a bend in 
the source spectrum to suppress low-energy particles in order to avoid 
too high an energy integral.)
However, $E^{-2.4}$ is quite close to the deduced spectrum required from SNR. 

\begin{figure}
\centering
\vspace{155pt}
\includegraphics{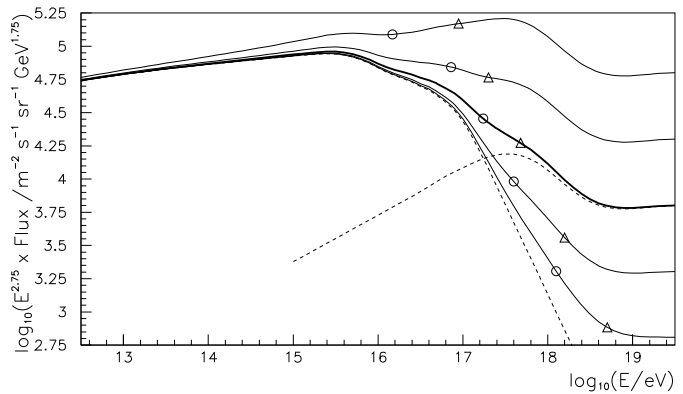}
\caption {To show that an almost imperceptible join between galactic and extragalactic
cosmic rays does not require a specially chosen relative strength of these two
components, the above plot shows the observed spectrum of cosmic rays (thick line),
and the spectra that would result if the supposed (proton-only) extragalactic part
(dashed line) is raised or lowered by a factor $\sqrt{10}$ or $10$, keeping the
Galactic component (dotted line on left side) unchanged.
Circles and triangles mark the
energies where 50\% and 80\% of the total flux is extragalactic.
Only at very low levels of extragalactic component does an ankle mark the transition,
as widely assumed.
\label{diffegal}
 }
\end{figure}
   (b) `EGAL H+He' is a ``primordial'' hydrogen-helium mixture, 
with source spectrum $E^{-2.2}$, close to the expected form $E^{-2.23}$ for 
acceleration at an ultrarelativistic shock (Achterberg et al. \cite{urslope}),
 and sources evolving as before.  
If $E_{max}$ is set at $5\times 10^{20}$ eV it lowers the proton 
component flux by essentially the same amount as in case (a).  
The proton part of the flux is shown by the line `p in HHe'.  
Here, the helium component becomes unusually important at a few EeV, 
because pair production starts at the same Lorentz $\gamma$ as for protons, 
which is at an energy 4 times higher.

    (c) Normal elemental composition at source.  
Allard et al (\cite{allard}) have discussed such a composition (as well as the pure 
proton case), which results in many other nuclei surviving to Earth at 
around $10^{19}$ eV, in addition to the helium seen in case (b), 
so the characteristic ``valley'' seen at the ankle is lost: see the
dotted line `EGALmixed'.  
The ankle would in this case indeed mark the point where an extended 
Galactic spectrum falls well below a much harder extragalactic flux. 
Their source spectrum has the form $E^{-2.3}$, and no evolution of source strength.

     Of these three extragalactic variants, the proton composition (a), 
advocated for some time by Berezinsky and associates, is probably the most 
physically plausible. Subtracting this flux from the observed total cosmic ray
flux requires that all other contributions to cosmic rays at Earth match the
curve marked ``Total\ --\ egalP'' in figure \ref{augere}, which is in excellent
agreement with the hatched band indicating the probable total Galactic cosmic ray
flux, assuming that the most massive nuclei (Si-Fe) in the KASCADE flux unfolding
(figure 2) do have a rigidity spectrum terminating like that of light elements,
so their flux falls away above $10^{17}$ eV as shown.

It may appear to be a remarkable accident that this steeply falling  
Galactic flux (presumed from SNRs) and the unrelated rapidly emerging 
extragalactic flux, add together to give a relatively smooth seamless total 
spectrum between the knee and the ankle.  The position of the extragalactic
flux emergence is governed by the threshold of the pair-production reactions between 
protons and CMBR photons, of course, and has not been adjusted to make a smooth
fit here.  
Figure \ref{diffegal} shows that one could alter the level of the extragalactic 
component considerably without providing a visible clue to the join point in the
overall spectrum.   The smoothness of the total spectrum can hide large bumps in 
individual components: indeed when the extragalactic part is increased 
tenfold (top curve), the usual knee due to the rigidity limit of Galactic accelerators
has become imperceptible!
So the appearance of a smooth total spectrum can easily completely hide major
disappearances and appearances of individual components.  
A smooth continuation from Galactic to extragalactic cosmic rays
is, surprisingly, quite natural.

In the case of Hillas's (\cite{hillasdsar}) H+He version of the 
extragalactic flux, there has to be a higher Galactic contribution at 0.1
to 0.4 EeV, which he tentatively attibuted to a high-speed part of the
type II SNs, following Bell and Lucek (\cite{belluc}).  These would have provided
a low-level extension to the rigidity spectrum beyond the normal $E_{max}$ of SNR
cosmic rays.  However, the present KASCADE analysis has not so far suggested any
such tail after the initial sharp drop at the knee.

An important point is that in both cases (a) and (b), the extragalactic 
component becomes a major part of the flux not far above $10^{17}$ eV, 
at which energy heavy nuclei still leak only slowly from the Galaxy, 
so very little anisotropy is to be expected at any energy above the knee, 
thus providing a surprising explanation of the lack of progress in the long 
search with larger and larger exposures for directional clues to the origin 
of cosmic rays.  
Composition (c) on the other hand produces a later transition to extragalactic 
particles, and a correspondingly greater Galactic component to be found
to fill the gap between the hatched region (end of the SNR component) and the
ankle.

The elemental composition just below $10^{18}$ eV will provide a test between 
the alternatives.  
Evidence from $x_{max}$ measurements from the Stereo Fly's Eye and Yakutsk 
gave no indication  of the emergence of a light component in this region
(e.g. see Allard et al. \cite{allard}), but HiRes on 
the contrary indicated a rapid change near here 
(Abu-Zayyad et al.\cite{hirescomp1}, Abbasi et al. \cite{hirescomp2}).  
If this version is confirmed, it supports the
pure proton model (a), and, presumably, injection in a region of very intense
photon flux.
Lateral distribution of shower particles has potentially equally good resolution 
in searching for a double-peaked composition (Fe + H).  
At present the Haverah Park data (Ave et al. \cite{hpcomp}) have suggested a modest 
proton component, as in model (b), but a re-analysis of other closely-spaced 
detector array results might be useful.

     The most important clue to the sources at very high energy should come 
from the long exposure of the Auger project, when several particles 
above $9 \times 10^{19}$ eV have been recorded: such particles cannot have
travelled more than about 50 Mpc, so magnetic deflections should not be large, 
and there should be few candidate sources within range in the directional error boxes.

\section{Some proposed cosmic-ray sources not reviewed above: connection with gamma 
ray bursts (GRB); cannonballs?\label{grbsec}}
This section has been added, in ``version 2'' of this paper, in response 
to questioners (but does not appear in the conference proceedings).  
The present review has focused on the attempt to understand most cosmic rays 
without ``new physics'', being strongly guided by theoretical expectations.
I have concentrated on what seems much the most probable 
scenario, in order to seek a quantitative understanding of the principal 
features of cosmic rays, omitting consideration of other sources which may 
make some contribution in particular energy domains.  
Other sources not discussed above, which may contribute at the few percent level,
include high-energy stellar winds, pulsars, magnetars, microquasars, shocks 
outside galactic discs and gamma-ray-burst phenomena.
(Of course, the specific sites within AGN environments for the extragalactic 
sources have not been examined either.)  
Thus pulsars/plerions appear to be sources of relativistic electrons, but 
probably at the few-percent level, as argued in section 1.
Only the particular case of the GRB-related phenomena will be taken up here,
to explain why they were not considered above as major contributors to
observed cosmic rays.

GRBs excited considerable interest in relation to particle acceleration as they
gave evidence of explosive events ejecting matter in narrow cones with Lorentz 
factors $\sim 100-1000$, and probably involving kinetic energy somewhat 
larger than that released in ordinary supernova explosions 
--- up to 30 f.o.e. ($=30 \times 10^{51}$ erg).  
The difficulty of finding a site where protons could be accelerated to 
$10^{20}$ eV (Hillas \cite{araa}) could be greatly alleviated if the 
accelerator were sited within, or comprised, a jet (or ``plasmoid'') moving 
with a large Lorentz factor, $\Gamma_{jet}$, as the maximum energy of particles
released into the environment from an accelerator of given $R\times B$ could 
be increased by a factor $\Gamma_{jet}$. 
(An additional helpful factor $\Gamma$ can appear when considering the
limit set by a small value of $B$ ahead of the shock if the inflow to the shock
is ultrarelativistic.)
AGN jets are favoured sites for acceleration of the most energetic cosmic rays 
for this reason of course (together with the large available kinetic energy), 
but GRBs might have even higher $\Gamma_{jet}$.  The ``long GRBs'' are the 
relevant ones, thought to be most probably a variety of supernova explosion 
that generates ultrarelativistic ejecta, probably arising from the collapse 
of a star of initial mass $>25 M_\odot$ to a black hole 
(the more massive type Ic SN). 

The reason why GRBs had not been considered here as a promising main 
source is that their summed power does 
not seem to compete with the sources discussed earlier.  
Woosley and Heger (\cite{woosgrb}) use the statistics of GRB detection to 
estimate that they represent less than 1\% of core-collapse SN, and interpret
them as resulting from an unusual form of collapse of massive stars of high 
spin rate.  
More quantitatively, one can ask whether GRBs could supply the energy in the
extragalactic cosmic rays described in the previous section.
Taking the preferred proton component (``EGAL p'' in figure \ref{augere}), the
local energy density in this $E^{-2.4}$ flux above 300 GeV is 
$2.4 \times 10^{-17} \rm{erg \ cm^{-3}}$.  (If we make use of gases moving with
Lorentz factors 300-1000 and then slowing, the output might effectively start 
at $\sim 300$ GeV.)
Today's local flux is equivalent to the output of sources of present-day strength
added over $\approx 2.8 \times 10^{10}$ yr, after integrating over the evolving
source strength (Porciani and Madau SFR2) and allowing for red shift.  
Hence the present-day
sources have to supply $\approx 2.5 \times 10^4 \rm{\ f.o.e.\ Gpc^{-3} yr^{-1}}$.
The local rate of GRBs, after correcting for beaming, is estimated by Guetta
et al. (\cite{grbrate}) as $\sim 33 \rm{\ Gpc^{-3} yr^{-1}}$; and taking their
average kinetic energy $\sim 20$ f.o.e. and a 1/6 acceleration efficiency, their
relativistic particle production would be $\sim 110 \rm{\ f.o.e. Gpc^{-3}yr^{-1}}$ 
--- too low by a factor $\sim 200$.
The alternative choice of the
``EGAL H+He'' flux in figure \ref{augere} reduces the energy deficit above
300 GeV to a factor $\sim 10$.
This is very different from the conclusion of Vietri, De Marco and Guetta
(\cite{vietuhe}) that the power required to supply the most energetic cosmic
rays is less than the GRB output.  The main reason for the difference is that
Vietri et al. consider only the energy needed to supply cosmic rays from 
$10^{17}$ eV upwards, arguing that particles below $10^{17}$ eV cannot escape 
from the front of the ultrarelativistic GRB.  
However, the Fermi acceleration generates a full spectrum from injected energy
upwards, behind the shock, and the bulk of the energy provided resides in the
particles below $10^{17}$ eV (observed energy).  
This is the major part of the energy budget.  
(Although it is now suspected that weak bursts (X-ray flashes) may be more
frequent than standard GRBs by an order of magnitude (e.g. Soderberg et al., 
\cite{xrfrate}), these put little energy into only mildly relativistic shocks.)
All this deserves a fuller treatment, of course.

In particular, a supernova-related model is unattractive for explaining 
extragalactic cosmic rays, as our galaxy would dominate the local flux, on 
geometrical grounds (Hillas \cite{hillascris}) if the accelerators 
were found in all galaxies.  Of course, the local flux arising from infrequent 
outbursts could suffer large fluctuations about the average, accentuated if 
the output were strongly beamed,
but Galactic magnetic fields should provide some
smoothing for moderately heavy nuclei before the ankle, if the outbursts 
were not much below 1\% of the regular supernova rate. 
This is why one looks for exceptionally powerful sources not found in normal
galaxies at the highest energies.

Quite unlike Woosley and Heger, Dar and De R\'{u}jula have presented the 
hypothesis that essentially all core-collapse SN, rather than $<1\%$, produce 
the ultrarelativistic ejections that generate GRB phenomena, but which would 
only be seen directly in an exceedingly narrow observing cone.  
As an unusual basis for explaining GRB and afterglow phenomena, these authors 
propose that most core collapse supernovae eject several ``cannonballs'' of 
plasma over a few days, having Lorentz factors 
clustered near 1000, carrying in total an energy $\approx 2$ f.o.e..  
The authors have gone on to estimate (most fully in Dar and de R\'{u}jula
\cite{darderu}) that these objects
would also naturally account for the generation of virtually
all cosmic rays as a consequence of  their motion through the interstellar gas.
However, I can give no credence to this model.
Firstly, the vital properties of the cannonball (CB) are its Lorentz factor,
$\Gamma$, which determines the energy of the particles it emits, and its
transverse radius, $R$, which determines the rate at which particles are
swept up and the CB slows down.  
A quite false effect is employed (Dado, Dar and De R\'{u}jula \cite{ddd}) 
in calculating that the expansion of $R$ is
quickly slowed down from $\sim 10^5 \rm{m\ s^{-1}}$ (when emitted from an 
accreting neutron star) and that $R$ almost stabilises at $10^{12}$m, expanding
thereafter at $\sim \rm{m\ s^{-1}}$ (in the standard stellar frame).  
The swept-up particles, which have a
very high individual energy $\Gamma m_p c^2$ in the CB reference frame, 
and which are supposed to diffuse out of the CB, are said to exert an
inward pressure, opposing lateral expansion, supposedly because of momentum
reaction when they leave, whereas the opposite is true: near the edges, where
the net particle flow is outwards, a net {\it outward} force (pressure gradient)
would be exerted by the diffusing particles on the CB material. 
Secondly, the authors suppose that particles can simply leave the CB's surface,
without Fermi ``bouncing'', whereas Achterberg et al. (\cite{urslope})
show that particles entering a plasma advancing with ultrarelativistic speed,
and scattered back out of it, cannot escape: even a small
external magnetic field retards them sufficiently that they are recaptured,
and a Fermi acceleration process is set up that dominates the spectrum.
(If $B$ were as
small as the normal Galactic field $\sim{2}\rm{\mu G}$ they would still be
recaptured within the CB radius until their rigidity exceeded $\sim 10^{17}$ V.)
The CB authors do wish to use Fermi acceleration to get particles to
$10^{21}$ V, however, but seem to misinterpret the reports of Frederiksen 
et al. \cite{fredurelmagf}, \cite{fredelec}, as showing that
relevant acceleration of ions will occur entirely behind the shock, so they 
considered motion entirely there (with a field strength $\sim \rm{gauss}$).
Frederiksen et al., did note, though, that the magnetic field in the
external medium was required for significant energy gain by ions.

\section{Conclusions}

Cosmic ray physics is perhaps becoming less exciting for seekers of the exotic.  
There is as yet no
necessity for new physics at the highest energies; and we may be approaching
the situation where the mysteriously bland spectrum between the knee and the ankle 
is resolved into the sum of perhaps only two major kinds of source.
If this interpretation is correct (notably the proton source model of Berezinsky 
et al.), the transition from Galactic to extragalactic cosmic rays has
occurred at a much lower energy than was usually believed, and has left
virtually no obvious sign in the flux level at the join point.
Nevertheless, a rapid change in shower characteristics should occur here, and
measurements of double-peaked distributions may be possible.
The detection of a class of air showers of very uniform structure near $10^{17}$ eV,
attributale to the most energetic Galactic cosmic rays, and rapidly diminishing in
proportion as energy rises, would help to establish clearly whether  proton-only
acceleration occurs at the highest energies. 

Such an interpretation, and the demonstration of sharp sub-knees in the
Galactic spectrum by the KASCADE experiment, have presented a lesson
that the simplicity of a smooth spectrum, close to a power law, can be very deceptive.
Cosmic-ray physicists should resist the temptation to read much into the position
of a ``join point'' between two straight lines drawn through flux data points.
The absence of obvious concavity in cosmic-ray proton spectra before the knee,
despite its prediction in diffusive shock acceleration, may be another instance
of many sub-spectra adding to give the appearance of a close approximation
to a power law in the total.

  A much better understanding of the knee as a consequence of SNR development
brings together theoretical and observational work very fruitfully.

The central part played in astronomy by detailed images (as well as spectra)
is exemplified by wonderful CHANDRA X-ray images of SNRs, and TeV detectors
are at the threshold of this capability..

\begin{acknowledgements}
I am particularly grateful for the stimulation of discussions with Tony Bell, 
Venya Berezinsky, Tom Gaisser, Todor Stanev and Gaurang Yodh while preparing this 
material, and for much assistance by Jamie Holder and other colleagues ---
also for comments and questions from several fellow cosmic-ray enthusiasts which
resulted in the addition of much material in section \ref{grbsec}.
\end{acknowledgements}

\end{document}